\documentclass[12pt,a4paper]{article}

\usepackage{epsf}
\usepackage{amsfonts}
\usepackage{graphicx}

\def\lesssim{{\buildrel < \over \sim}}

\begin{document} 

\title{Quantum Tests of the Foundations of General Relativity\footnote{To appear in {\it Class.\ Quant.\ Grav.}}
} 

\author{Claus L\"ammerzahl\thanks{e-mail: Claus@spock.physik.uni-konstanz.de} \\  
{\normalsize Laboratoire de Gravitation et Cosmologie Relativistes,} \\  
{\normalsize Universit\'e Pierre et Marie Curie, CNRS/URA 769, 75252 Paris Cedex 05, France} \\  
{\normalsize and } \\  
{\normalsize Fakult\"at f\"ur Physik der Universit\"at Konstanz, D - 78434 Konstanz, Germany}} 

\maketitle 

PACS--Nr. 04.80.Cc, 03.75.Dg, 03.65.-w


\begin{abstract}
A new test theory for describing tests of fundamental
principles of Special and General Relativity is presented.
Using a generalised Pauli equation which may be based on
a generalised Dirac equation, possible violations of local
Lorentz invariance and local position invariance on the
quantum level are described. On the quantum level there
are more possibilities to violate these principles than
on the classical level. The corresponding terms can
be tested by Hughes--Drever type experiments or by atom
interferometry. It is proposed that an atom interferometric test of 
Local Position Invariance will give a three order improvement of 
existing estimates.  
\end{abstract}


\section{Introduction} 

General Relativity (GR) is based on Einstein's Equivalence
Principle (EEP) which implies that gravitation is a metric theory.
EEP consists of the weak equivalence principle (WEP), local
Lorentz invariance (LLI) and local position invariance (LPI);
for a review see \cite{Will93,Will95}.
All these notions essentially rely on point particles and paths.
Therefore GR is primarily applicable to the motion of satellites,
planets and light rays.
However, since quantum matter possesses more degrees of freedom
and is always spread out at least over a certain space-like region,
a theory of gravity
at the quantum level may be different from GR at the classical
level and therefore should be founded on principles using
quantum mechanics only.
It may be possible, for example, that gravity on the quantum level has to
be described by more fields than the metrical field.
In addition, for doing tests of the foundations of GR
on the quantum level, one
should take into account that there may be more possibilities
to break LLI and LPI on the quantum level than on the classical level.

Therefore, in order to test gravitational theories at the
quantum level it is necessary to use a test theory which is
based solely on quantum notions.
We do not transcribe classical notions into the quantum
domain, e.g. introduce a PPN metric or the corresponding
$\gamma$--matrices into the Klein--Gordon or Dirac equation.
One should allow the most general interaction on the quantum
level.
We base our test theory on a generalised Pauli equation (GPE)
which on the one hand is general enough to allow violations of
WEP, LLI, and LPI, and on the other hand still obeys
fundamental quantum principles.
An important point is that we include spin which is necessary
since there are no massive particles without spin.

An important point of our considerations is that in order to arrive at our test theory we do not use any geometrical notion. 
This is necessary from a logical point of view since we want to reach the result, that only under certain circumstances (experimental results), gravitation can be described by means of a space--time geometry. 
In addition, we do not use a Lagrangian formalism. 

Our test theory is a test theory for both special
relativity {\it and} GR together.
In contrast to the test theory of Robertson \cite{Robertson49}
and  Mansouri--Sexl \cite{MansouriSexl77}, and also their gravitationally modified form \cite{TourrencMellitiBosredon96}, our theory does
not rely on a selected frame of reference.
It is the {\it dynamics} of the quantum field which is responsible for 
whether LLI is violated or not.
The coupling to gravity is accomplished by introducing the
Newtonian potential in a most general way which may be particle dependent.

There are other test theories, like the $TH\epsilon\mu$
\cite{LightmanLee73,Will93} formalism, which allows a difference
in the velocity of light and the maximum speed of massive particles.
The main difference between this and our test theory, is that in our
case the violation of
the EEP is due to the structure of the dynamical equation of one
matter field, and not due to an `anomalous' interaction
between matter and the electromagnetic field.
In this sense our test theory is not an alternative to the $TH\epsilon\mu$--formalism, but instead a theory
of one single multicomponent matter field.
The structure of the coupling of the GPE (or of the generalised
Dirac equation on which we base our GPE) to the electromagnetic field is another question. It may indeed happen that both fields, the GPE and the Maxwell field,
are dynamically incompatible in the sense that they possess to
two different maximal velocities of
propagation which of course leads to certain observable consequences. 
It is the $TH\epsilon\mu$ formalism which describes the experimental
consequences due to this
difference in the maximum speed in, for example, atomic systems.
Here we skip this point and assume the usual minimal
coupling to the electromagnetic field obeying the usual
Maxwell equations. 

Best estimates for parameters characterising the violation of LLI
and LPI are given by Hughes--Drever type experiments and by the
Vessot--Levine rocket red shift experiment (for a recent review
see \cite{Will95}). 
LPI and LLI violating parameters are constraint by $10^{-4}$ and
$10^{-30}$, respectively. 
Here we want to indicate that in some cases more precise tests are possibleusing atom beam interferometers of the Kasevich--Chu type \cite{KasevichChu91,KasevichChu92}. 
If one carries through atom beam interferometric tests in the field of the earth with two different types of atoms, then the accuracy of these devices may test the validity of the WEP in the quantum domain with an accuracy $10^{-8}$ (it is announced that these interferometers may be improved in the near future) and may especially restrict the LPI violating parameter to an accuracy of about $10^{-7}$ which will be an imoprovement of the present estimate by three orders. 
Hughes--Drever type experimenbts are already carried through, and we use a reanalysis of these experiments for calculating constraints for the parameters of our test theory. 
In addition, since all experiments made on earth are in fact done in a noninertial frame moving in a gravitational field, each high precision experiment can be used as a test of the underlying space--time structure. 

In the following we treat the GPE which we derive from
a non--relativistic limit of a generalised Dirac equation (GDE).
Then we 
calculate some experiments, namely atom beam interferometry
\cite{Borde89},
and the modification of energy levels of atoms, and give
estimates for the various parameters describing
LLI-- and LPI--violation.

\section{The generalised Pauli equation}

\subsection{The generalised Dirac equation}

We base our non--relativistic test theory on a generalised Dirac equation (see, for example, 
\cite{AL93,BleyerLiebscher95})
\begin{equation}
i \partial_t\varphi = - i c (\widetilde\alpha^i \nabla_i
+ i \Gamma) \varphi + m c^2 \widetilde\beta \varphi + e \phi \varphi \label{GDE}
\end{equation}
($i, j = 1, 2, 3$) where $\varphi$ is a complex 4--component field
(it is possible to carry through the following consideration for higher component fields too; however, we want to restrict to the physically most important case which is connected to spin--${1\over 2}$--fields). 
All matrices are complex $4\times 4$-matrices and $\widetilde\alpha^i$
and $\widetilde\beta$ obey $(\widetilde\alpha^i)^+ = \widetilde\alpha^i$,
$\widetilde\beta^+ = \widetilde\beta$, and $\Gamma^+ =
\Gamma + i \partial_i \widetilde\alpha^i$.
However, they are not assumed to fulfill a Clifford algebra.

Here we introduced the coupling to the Maxwell field in the usual manner namely through minimal coupling $\nabla_i = \partial_i - {{i e}\over{c}} A_i$. 
We also take the usual form of the Maxwell field as granted (for the experimental status of the electromagnetic field to obey EEP compare \cite{HauganKauffmann95}). 
It is of course also possible to couple the Maxwell field in an anomalous way to the GDE, analogous to the $TH\epsilon\mu$--formalism. 
However, since any modification of this kind will result in corrections of the same structure as those which we will derive below, we will not take anomalous couplings to the Maxwell field into account. 

We also introduced a $c$ which has the dimension of a velocity. 
This velocity can be introduced by considering the null cones which the GDE defines: $c$ is the maximum speed of propagation (from the physical point of view it is approximately the velocity of light, because any deviation from SRT, if there is any, is small). 
The purpose of this velocity is twofold: First, it makes the coefficient in front of the spatial derivative dimensionless (what is necessary in order to connect $\widetilde\alpha^i$ with space--time geometry), and second, it will be used later as ordering parameter in a Foldy--Wouthuysen transformation leading to the non--relativistic limit of the GDE.

The splitting between the matrices $c\Gamma$ and $m c^2 \widetilde\beta$ is defined by means of a WKB approximation (compare \cite{La90}). 
While $m c^2 \widetilde\beta$ is the ``mass''--tensor which appears in the lowest order of approximation, $\Gamma$ influences the first order only. 
Both matrices have the dimension of ${\hbox{length}}^{-1}$. 
In order to extract from the ``mass''--tensor a dimensionless matrix possessing a geometrical meaning, we introduced a parameter $m$ (so that $m c^2$ has the dimension ${\hbox{time}}^{-1}$) which can also be defined via the WKB approximation. 

Eqn.\ (\ref{GDE}) is general enough to describe violations of basic
principles of GR.
However, since (\ref{GDE}) is a symmetric hyperbolic system
very general principles of quantum mechanics are still fulfilled, namely
(i) the well--posedness of the Cauchy problem, (ii) the
superposition principle, (iii) finite propagation speed,
and (iv) a conservation law. 
Indeed, it has been shown that this GDE can be derived from these fundamental principles, see \cite{AL93} for a review. 

If we introduce the quantities 
\begin{eqnarray}
{4\over{g^{00}}} & := & \hbox{tr}\widetilde\beta^2 \label{g00} \\ 
{\hbox{\bf g}}^{0 i} & := & {{g^{0 i}}\over{g^{00}}} := {1\over 4} \hbox{tr}(\widetilde\alpha^i) \label{gij} \\ 
{\hbox{\bf g}}^{ij} & := & - {{g^{ij}}\over{g^{00}}} := {1\over 4} \hbox{tr}(\widetilde\alpha^i \widetilde\alpha^j) - 2 {\hbox{\bf g}}^{0 i} {\hbox{\bf g}}^{0 j} \label{bfgmunu} 
\end{eqnarray}
then the matrices $\widetilde\alpha^i$ and $\widetilde\beta$ fulfill
\begin{eqnarray}
\widetilde\alpha^{(i} \widetilde\alpha^{j)} - {\hbox{\bf g}}^{ij} 1 
-  2 {\hbox{\bf g}}^{0 (i}\widetilde\alpha^{j)} & = & X^{ij} \\  
\widetilde\alpha^i \widetilde\beta + \widetilde\beta \widetilde\alpha^i -  2 {\hbox{\bf g}}^{0 i} \widetilde\beta & = & 2 X^i \\ 
\widetilde\beta^2 - {1\over{g^{00}}} & = & X
\end{eqnarray}
where the deviation from the usual Clifford algebra is described by the matrices $X$, $X^i$, and $X^{ij}$. (In the case $X = 0$, $X^i = 0$, and $X^{ij} = 0$ one can represent $\alpha^i = (\gamma^0)^{-1} \gamma^i$ and $\beta = (\gamma^0)^{-1}$ with matrices $\gamma^\mu$ fulfilling $\gamma^{(\mu} \gamma^{\nu)} = g^{\mu\nu}$, $\mu,\nu = 0, \ldots 3$. 
Even in the case that the $X$--matrices do not vanish it can be shown \cite{La90} that the matrices $\widetilde\alpha^i$ and $\widetilde\beta$ fulfill a generalised Clifford algebra.) 
If the matrices $\widetilde\alpha^i$ and $\widetilde\beta$ do not fulfill the usual Clifford algebra then the characteristic surfaces (null cones) and the mass shells (see Figure) of the generalised Dirac equation split and do not longer coincide with the usual light cones and mass shells.
It is obvious that in these cases LLI is violated.
This has also been discussed in \cite{Liebscher85} (see also
\cite{BleyerLiebscher95} and \cite{ABL93} and references therein).

A velocity $c$ can also be introduced by means of the quadratic form based on the tensor $g^{\mu\nu}$ defined by (\ref{g00}--\ref{bfgmunu}), which can be chosen to possess the signatur 2 compatible with the required propagation properties of the field $\varphi$. 

\begin{figure}[t]
\begin{center}\framebox{\begin{minipage}[t]{16.5cm}\includegraphics*[scale=0.4]{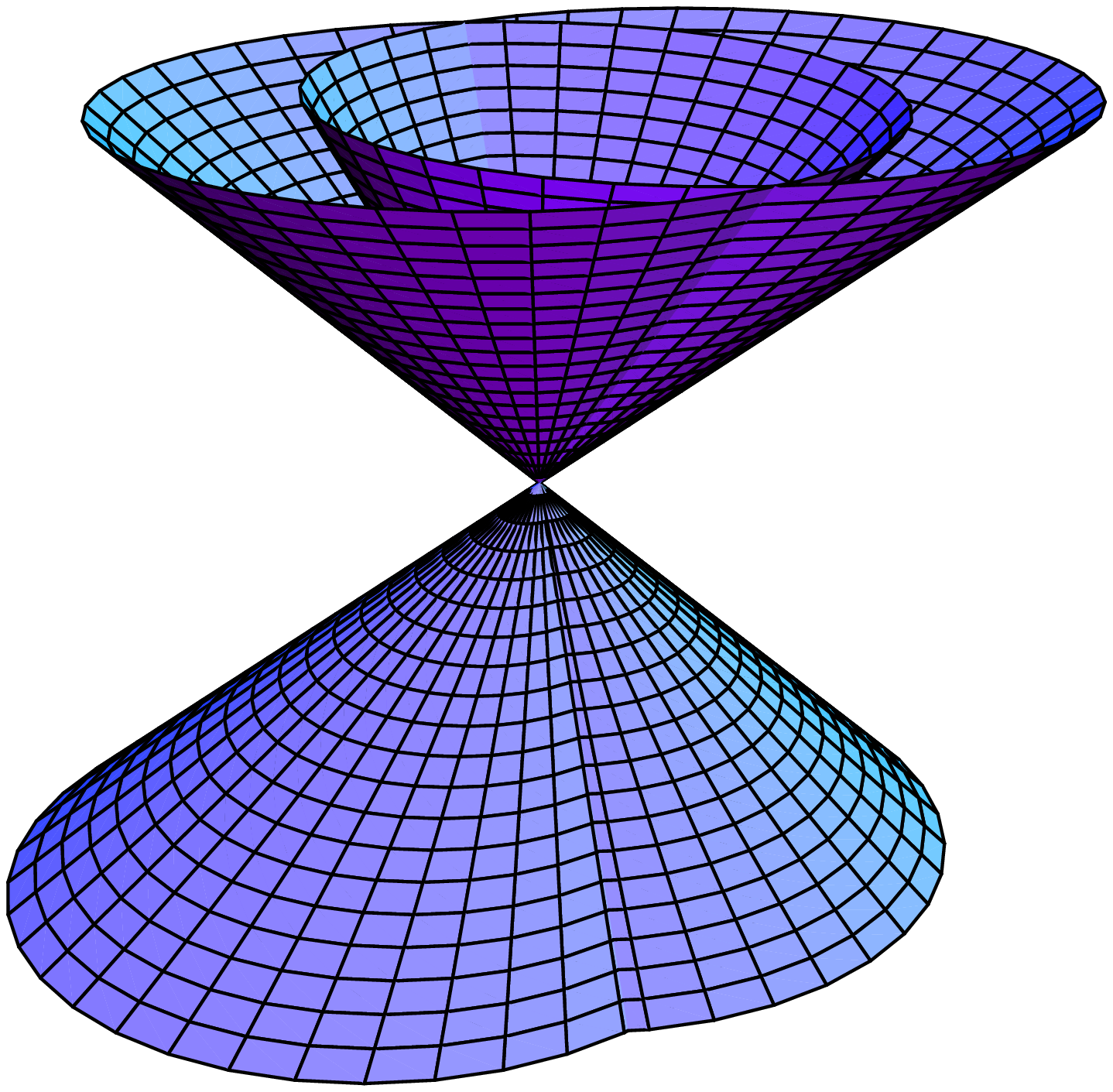} \includegraphics*[scale=0.4]{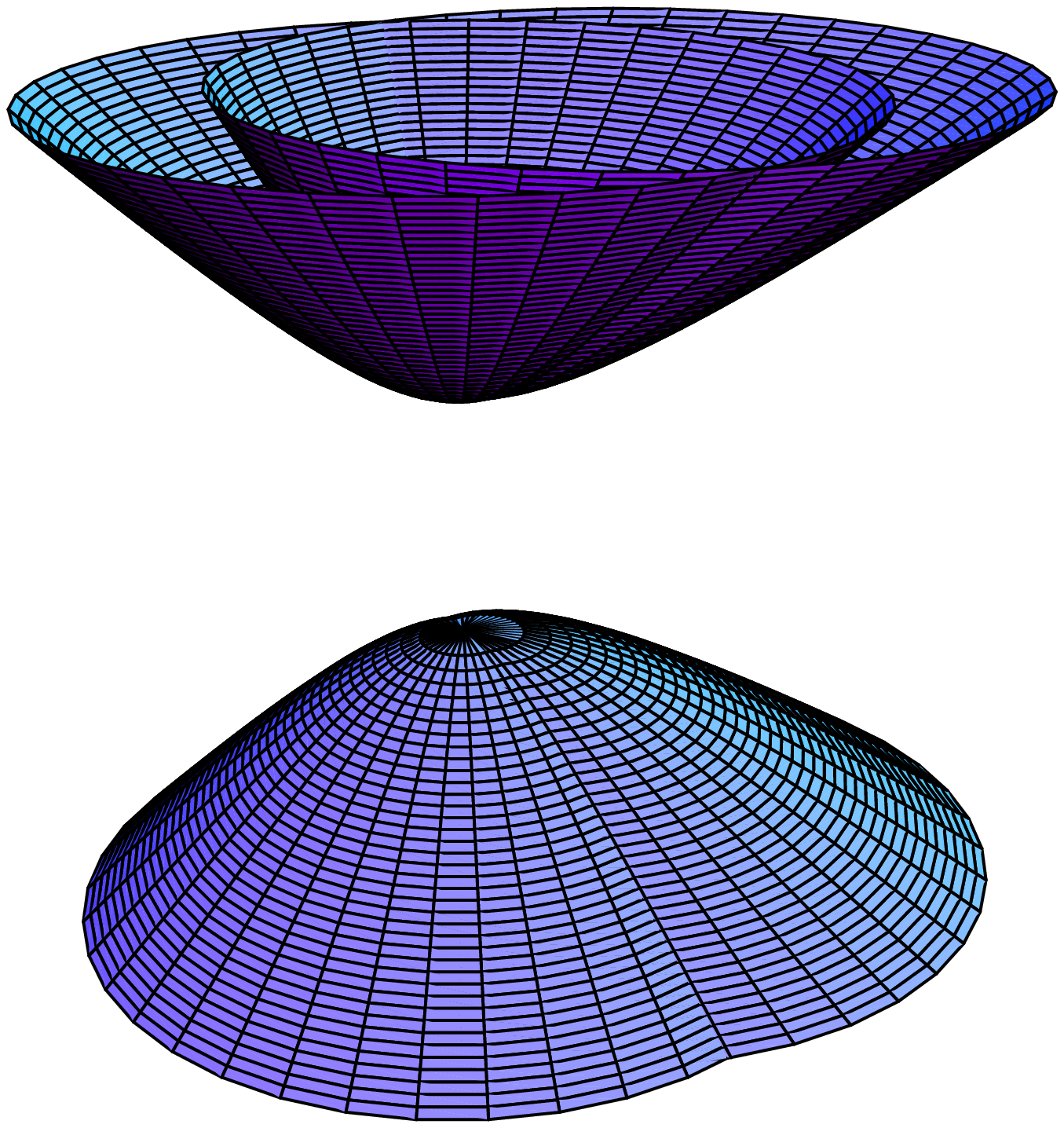}
\end{minipage}}\end{center} 
\end{figure}

\subsection{Non--relativistic approximation: The generalised Pauli equation}

While the Dirac equation is an equation for four spinorial components, a non-relativistic limit, the Pauli equation, has two components only. 
We have to eliminate two components which are small in the physical situation under consideration. 
The matrix which serves as tool for distiguishing the upper and lower (that is, the large and small) components is, as usual, $\beta = \left(\matrix{1 & 0 \cr 0 & -1 \cr}\right)$. 
We also introduce as usual even and odd operators $\cal E$ and $\cal O$ which fulfill $\beta {\cal E} = {\cal E} \beta$ and $\beta {\cal O} = - {\cal O} \beta$. 

First we split the matrices $\widetilde\alpha^i$ and $\Gamma$ into an even and odd part: $\widetilde\alpha^i = \widetilde\alpha^i_e + \widetilde\alpha^i_o$ and
$\Gamma = \Gamma_e + \Gamma_o$. 
We have $\hbox{dim}(\widetilde\alpha^i) = \hbox{dim}(\widetilde\beta) = 1$ and $\hbox{dim}(\Gamma) = \hbox{dim}(\partial_i)$. 
Then 
\begin{eqnarray}
i \partial_t\varphi & = & - i c (\widetilde\alpha^i_e + \widetilde\alpha^i_o) \nabla_i \varphi + c \Gamma \varphi + (\phi + m c^2 \widetilde\beta)\varphi \nonumber\\
& = & \beta m c^2 + {\cal E} + {\cal O}
\end{eqnarray}
with
\begin{eqnarray}
{\cal O} & = & - i c \widetilde\alpha^i_o \nabla_i + c \Gamma_o + m c^2 \widetilde\beta_o \\ 
{\cal E} & = & - i c \widetilde\alpha^i_e \nabla_i + c \Gamma_e + m c^2 \widetilde\beta_e - m c^2 \beta
\end{eqnarray}
Performing a Foldy-Wouthuysen-transformation (see, for example, \cite{BjorkenDrell64}) with $\varphi^\prime = U \varphi$, $U = e^{i S}$, $S = - {i\over{2 m}} \beta {\cal O}$ with $S^+ = S$ (since $H$ as well as $\beta m c^2$ is hermitean we also know that ${\cal E}$ and ${\cal O}$ are hermitean) we get as resulting Hamiltonian 
\begin{eqnarray}
H^\prime\varphi^\prime & = & \beta \left(m c^2 + {{{\cal O}^2}\over{2 m c^2}} - {{{\cal O}^4}\over{8 m^3 c^6}}\right)\varphi^\prime + {\cal E}\varphi^\prime - {1\over{8 m^2 c^4}} \left[{\cal O}, [{\cal O}, {\cal E}]\right]\varphi^\prime - {i\over{8 m^2 c^4}} [{\cal O}, \dot{\cal O}]\varphi^\prime \nonumber\\
& = & \beta m c^2 \varphi^\prime - \beta {1\over{2 m}} \widetilde\alpha^{(i}_o \widetilde\alpha^{i)}_o \nabla_i \nabla_j \varphi^\prime \nonumber\\
& & - \Biggl[\beta {1\over{2 m}} (\widetilde\alpha^i_o\nabla_i \widetilde\alpha^j_o + i \{\Gamma_o, \widetilde\alpha^j_o\} + i m c \{\widetilde\beta_o, \widetilde\alpha^j_o\}) + i c \widetilde\alpha^j_e\Biggr] \nabla_j\varphi^\prime \nonumber\\
& & + \Biggl[\beta \Bigl({1\over{2 m}} (\Gamma_o)^2 + {1\over 2} m c^2 (\widetilde\beta_o)^2 + {c\over 2} \{\Gamma_o, \widetilde\beta_o\} - {i\over{2 m}} \widetilde\alpha^i_o \nabla_i(\Gamma_o + m c \widetilde\beta_o)\Bigr) \nonumber\\
& & \qquad + c \bar M_e^1 + m c^2 (\widetilde\beta_e - \beta) + \phi + {i\over{2 m}}\widetilde\alpha^{[i}_o \widetilde\alpha^{j]}_o H_{ij}\Biggr]\varphi^\prime + \hbox{relativistic terms} \label{GPE1}
\end{eqnarray}
with $H_{ij} = \partial_i A_j - \partial_j A_i$ and where we used a quasi--Newtonian coordinate system. 
With `relativistic terms' we mean such terms which contain a $c$ in the denominator. 
This Hamilton operator is of course hermitian since the operators $\cal E$ and $\cal O$ are hermitian. 

We now make some specifications of the coefficients appearing in (\ref{GPE1}) in order to arrive at a physically interpretable generalised Pauli equation. 

\begin{enumerate} 

\item To start with, we make an ansatz concerning the introduction of the gravitational field. 
First, the gravitational field may consist of all coefficients appearing in (\ref{GPE1}). 
However, in the non--relativistic regime the main part of the gravitational field is given by the Newtonian potential $U(x)$ depending on the mass density $\rho(x)$ of the surrounding matter distribution. 
In order to allow an anomalous coupling to the gravitational field the origin of which are mass distributions, we also take gravitational field tensors $U^{ij}(x)$ into account (see e.g \cite{Will93}). 
There may also exist gravitational fields the origin of which are not mass distributions, but for example  spin--sources which is the case in Einstein--Cartan theories, see \cite{HehlHeydeKerlick76}. 
Since these gravitational fields are expected to be very small we do not specify any spatial dependence of these fields: all gravitational fields which have their origin in other sources than in mass distributions are treated as constant. 
Therefore we assume that the $x$--dependence of the coefficients appearing in the above non--relativistic limit is only mediated by the Newtonian potential and the corresponding gravitational potential tensor \cite{Haugan79,Will93}. 
The remaining constant parts may be responsible for violations of LLI. 
Explicitly
\begin{eqnarray}
\widetilde\beta(x) & = & {\buildrel 0 \over {\widetilde\beta}} + {\buildrel 1 \over {\widetilde\beta}}{1\over{c^2}} \left(U(x) + {{\delta m^\prime_{\hbox{\scriptsize P} kl}}\over m} U^{kl}(x)\right) \label{widetildebeta} \\ 
\widetilde\alpha^i(x) & = & {\buildrel 0 \over {\widetilde\alpha}}{}^i + {\buildrel 1 \over {\widetilde\alpha}}{}^i {1\over{c^2}} \left(U(x) + {{\delta m^\prime_{\hbox{\scriptsize P} kl}}\over m} U^{kl}(x)\right) \\ 
\Gamma(x) & = & {\buildrel 0 \over \Gamma} + {\buildrel 1 \over \Gamma} {1\over{c^2}} \left(U(x) + {{\delta m^\prime_{\hbox{\scriptsize P} kl}}\over m}U^{kl}(x)\right) + {\buildrel 1 \over \Gamma}{}^i {1\over{c^2}} \left(\partial_i U(x) + {{\delta m^\prime_{\hbox{\scriptsize P} kl}}\over m} \partial_i U^{kl}(x)\right) 
\end{eqnarray}
where all coefficients ${\buildrel 0 \over {\widetilde\beta}}$, ${\buildrel 1 \over {\widetilde\beta}}$, ${\buildrel 0 \over {\widetilde\alpha}}{}^i$, ${\buildrel 1 \over {\widetilde\alpha}}{}^i$, ${\buildrel 0 \over \Gamma}$, ${\buildrel 1 \over \Gamma}$, and ${\buildrel 1 \over \Gamma}{}^i$ are constant matrices. 
While ${\buildrel 0 \over {\widetilde\beta}}$ and ${\buildrel 0 \over {\widetilde\alpha}}{}^i$ may be responsible for a violation of LLI, that is, for a violation of Special Relativity in vacuum, the other matrices may lead to $U$--induced violations of LLI and LPI. 

In other words: after the specification of the dependence of the matrices $\widetilde\beta$, $\widetilde\alpha^i$, and $\Gamma$  from the potential $U$ and $U^{ij}$ the remaining matrices ${\buildrel 0 \over {\widetilde\beta}}$, ${\buildrel 0 \over {\widetilde\alpha}}{}^{\hat\mu}$, and ${\buildrel 0 \over \Gamma}$  may also depend on $x$ in a way not mediated by $U$. 
However, since we are going to describe experiments which are performed on a small scale only (atom beam interferometry, Huges--Drever experiments) any $x$--depencence which is not due to the Newtonian potential can be neglected and be replaced by the actual value of these matrices at the position of the experiment. 
Therefore, the above dependence on $x$ is the only position--dependence of the corresponding matrices, that is $\widetilde\beta(x) = \widetilde\beta(U(x))$, and similarly for the matrices $\widetilde\alpha^{\hat\mu}(x)$ and $\Gamma(x)$. 

Although we know from experience that $U$ should be the Newtonian potential, at this stage of reasoning we do not need this specification so that at this place we take $U$ and $U^{ij}$ to be some unknown scalar and tensorial gravitational potentials. 
It is only at the last step where one makes a comparison with experiments that $U$ is identified with the Newtonian potential and $U^{ij}$ with the gravitational potential tensor. 

The dependence on $U^{ij}$ can be described by any parameter; we have chosen just for convenience and for staying in contact with the usual notation the combination $\delta m^\prime_{\hbox{\scriptsize P} ij}/m$. 

The only remaining $x$-dependent terms are the gravitational potentials $U$, $U^{ij}$, the electromagnetic potentials $\phi$, $A_i$, and the term $H_{ij}$. 

\item Each anomalous term which does not vanish for $U = 0$ describes a possible LLI-violation. 
Consequently, all terms which remain after setting $U = 0$, are small. 
Therefore we can neglect the square $({\buildrel 0 \over \Gamma}_o)^2$.
The $U$-dependent terms indicate possible LPI-violations. 

\item We make connection to the case of vanishing gravitational potential and non-rotating frame of reference: We take 
\begin{equation}
{\buildrel 0 \over {\hbox{\bf g}}}{}^{ij} = \delta^{ij}, \qquad {\buildrel 0 \over {\hbox{\bf g}}}{}^{0i} = 0, \qquad {\buildrel 0 \over g}{}^{00} = 1 
\end{equation}
and assume that these relations hold for all particles (neutrons, electrons, protons, etc). 

\item We can further simplify the form of $\widetilde\beta$ of (\ref{widetildebeta}) by choosing a representation in which ${\buildrel 0 \over {\widetilde\beta}}$ is even: ${\buildrel 0 \over {\widetilde\beta}}_o = 0$. 

\item We restrict to the `large' component using the projection operator $P = {1\over 2}(1 + \beta)$.
By doing so we introduce the following terms 
\begin{eqnarray}
P {\buildrel 0 \over {\widetilde\alpha}}{}^{(i}_o {\buildrel 0 \over {\widetilde\alpha}}{}^{j)}_o & =: & \delta^{ij} + {{\delta m_{\hbox{\scriptsize I}}^{ij}}\over m} + {{\delta \bar m_{\hbox{\scriptsize I} k}^{ij}}\over m} \sigma^k \label{anomalinert}\\   
P {\buildrel 0 \over {\widetilde\alpha}}{}^i_e & =: & A^i + A^i_j \sigma^j \\  
P \{{\buildrel 0 \over \Gamma}_o, {\buildrel 0 \over {\widetilde\alpha}}{}^i_o\}  & =: & 2 (a^i + a^i_j \sigma^j) \\ 
P {\buildrel 0 \over {\Gamma_e}} & =: & T + T_j \sigma^j \\
P {\buildrel 1 \over {\widetilde\beta}}_e & =: & 1 + d + C_j \sigma^j \\  
P({\buildrel 0 \over {\widetilde\beta}}_e - \beta) & =: & B + B_j \sigma^j \\  
P{\buildrel 0 \over {\widetilde\alpha}}{}^{[i}_o {\buildrel 0 \over {\widetilde\alpha}}{}^{j]}_o & =: & K^{ij} + (\epsilon^{ij}_{\phantom{ij}k} + K^{ij}_k) \sigma^k 
\end{eqnarray}
where $\sigma^i$ are the usual Pauli matrices. 
The anomalous inertial mass tensor in (\ref{anomalinert}) consists in two parts: a scalar part and a spin part. 
The hermiticity of $\widetilde\alpha^i$ ensures that $\delta m_{\hbox{\scriptsize I}}^{ij}$ as well as $\delta \bar m_{\hbox{\scriptsize I} k}^{ij}$ are real tensors. 
Note that $\delta m_{\hbox{\scriptsize I}}^{ij}$ and $\delta\bar m_{\hbox{\scriptsize I}k}^{ij}$ depends on $\widetilde\alpha^{\hat\mu}$ only, $\widetilde\beta$ has no influence on the kinetic term. 
We can absorb $d$ into the anomalous gravitational mass tensor giving $\delta m_{\hbox{\scriptsize P} ij}$.

For the usual Clifford algebra we have vanishing parameters $\delta m_{\hbox{\scriptsize I}}^{ij}$, $\delta \bar m_{\hbox{\scriptsize I} k}^{ij}$, $A^i$, $A^i_j$, $d$, $C_j$, $B$, $B_j$, $K^{ij}$, and $K^{ij}_k$. 
If any one of these parameters does not vanish, LLI is violated. 
The anomalous parameters which are connected with ${\buildrel 0 \over \Gamma}$ indicate a violation of LPI. 

If space--time is endowed with a hypothetical torsion then the usual Dirac equation minimally coupled to metric and torsion gives rise to the quantities $a^i_j$ and $T_j$. 
The latter is the space part of the axial torsion vector, and the first is related to the corresponding time component.

\item We transform away the term $\left(A^i + {1\over m} a^i\right) \nabla_i \varphi^\prime$ by means of $\varphi^\prime = e^{- i m c \delta_{ij} (A^i + {1\over m} a^i) x^j} \widehat \varphi$. 
The resulting constant scalar parts appearing in the last term of (\ref{GPE1}) can be removed by an appropriate transformation. 

\end{enumerate}
 
We arrive at
\begin{eqnarray}
H^\prime \widehat\varphi^\prime 
& = & - {1\over{2 m}} \left(\delta^{ij} + {{\delta m_{\hbox{\scriptsize I}}^{ij}}\over m} + {{\delta \bar m_{\hbox{\scriptsize I} k}^{ij}}\over m} \sigma^k\right) \nabla_i \nabla_j \widehat\varphi^\prime - \left({1\over m} a^i_j + c A^i_j\right) \sigma^j i \nabla_i \widehat\varphi^\prime  \nonumber\\
& & + \Biggl[e \phi(x) + {e\over{2 m}} H_i(x) (K^i + (\epsilon^{ij}_{\phantom{ij}k} + K^i_k) \sigma^k) \nonumber\\
& & + (m c^2 B_i + c T_i) \sigma^i + (1 + C_i \sigma^i) m U(x) + \delta m_{\hbox{\scriptsize P} ij} U^{ij}(x)\Biggr] \widehat\varphi^\prime \label{GPE}
\end{eqnarray}
where $\phi(x)$ is the electrostatic potential and $H_i = {1\over 2} \epsilon_{ijk} H_{jk}$ the magnetic field. 
This is the GPE we looked for. 
All terms but the $U$, $\phi$, $H_i$ and the $A_i$ which is part of the covariant derivative, are constant. 
The tensors $\delta m_{\hbox{\scriptsize I}}^{ij}$ and $\delta \bar m_{\hbox{\scriptsize I} k}^{ij}$ give spin dependent anomalous inertial mass tensors, $a^i_j$ and $A^i_j$ amount to a spin--momentum coupling, $m c^2 B_i$ may be considered as a spin--dependent ``rest mass'', 
the $T_i$ may be interpreted as or identified with the space--like part of an axial torsion vector, and $\delta m_{\hbox{\scriptsize P} ij}$ and $C_i$ are anomalous spin dependent passive gravitational mass tensors. 
$K^i$ and $K^i_k$ give anomalous modifications of the coupling of the magnetic field to the spin--${1\over 2}$ particle. 
Due to our systematic approach (\ref{GPE}) contains all possible anomalous interactions on the non--relativistic level. 
The GPE (\ref{GPE}) is a non--trivial generalisation
of Haugan's \cite{Haugan79} test theory for matter with spin.

A distinguished feature of (\ref{GPE}) is the anomalous coupling of spin to the Newtonian potential. 
Such couplings have been considered first by Hari Dass \cite{HDass77}. 
However, our coupling of the spin to the Newtonian potential is different from the coupling Hari Dass \cite{HDass77} introduced for spherically symmetric Newtonian gravitational fields.
While his couplings are of relativistic order on dimensional grounds, we obtained such a coupling by means of an additional anomalous property $C_i$ of the quantum field. 
Due to this new structure it is possible to have a coupling of the spin to the Newtonian potential at non--relativistic order. 

Note that there is no need and no possibility to introduce any $\hbar$. 
Indeed, also in the usual Schr\"odinger theory only the ratio $\hbar/m$ enters the equation of motion (see \cite{WeissYoungChu94}). 
In this sense our mass--like parameters are all to be understood in the sense of being the ratio of mass and $\hbar$. 
Our mass-like parameter has the dimension of $\hbox{time}/{\hbox{length}}^2$, and our Hamilton operator has the dimension $1/\hbox{time}$. 
It is no problem to introduce artificially an $\hbar$ so that the equations acquire the usual form and all parameters have the usual dimensions. 
Note also that $B_i$ is dimensionless. $T_i$ and $a^i_j$ have the dimension 1/length. 
In the following we will neglect the coupling of the magnetic field to anomalous terms. 

It should be emphasized that it is not possible to absorb the parameters $\left({1\over m} a^i_j + c A^i_j\right) \sigma^j$, $B_i$, and $T_i$ into the inertial or gravitational spin--dependent anomalous mass tensors. 
Indeed, if we perform a transformation $\varphi \rightarrow \varphi^\prime = S \varphi$, insert it into (\ref{GPE}), and demand that the resulting coefficient of the first derivative should vanish, then we get a transformation matrix $S$ which is linear in $x$. 
Such an $x$--dependence is not in accordance with a spherically symmetric Newtonian potential, for example. 
In addition, assuming a time--dependent transformation necessarily leads to time--dependent coefficients, which makes a stationary problem non--stationary thus changing the structure of (\ref{GPE}) dramatically. 

Since (\ref{GPE}) can be inferred from the very general GDE (\ref{GDE}) all anomalous terms in (\ref{GPE}) are derived in systematic manner. 
These are the most general anomalous terms on the
non--relativistic level which can be derived from a GDE which is the most general equation obeying the very general basic principles listed above.
The anomalous terms are necessarily connected with that parts
of the matrices $\widetilde\alpha^i$, $\widetilde\beta$, and ${\buildrel 0 \over \Gamma}$ which are responsible for a possible violation of LLI and LPI. 
A violation of LLI is possible even if the gravitational potentials are turned off. 

To sum up: we base our {\it quantum test theory} on a GPE (\ref{GPE}).
$m$ is the usual scalar mass, $\phi$ and $A_i$ the scalar
and vector electromagnetic potential and $H_i$ the magnetic part of the
electromagnetic field.
Thereby we assume that the electromagnetic interaction
has the usual form, that is, there are no EEP--violating
effects due to electromagnetism.
$\delta m_{\hbox{\scriptsize I}}^{ij}$ and
$\delta\bar m_{\hbox{\scriptsize I} k}^{ij}$ are
anomalous inertial mass tensors where the latter
is connected with the spin of the quantum system, and
$\delta m_{\hbox{\scriptsize P}ij}$ is the passive
gravitational anomalous mass tensor.
We introduced in addition a gravitational potential tensor
$U^{ij}$ with $\delta_{ij} U^{ij} = U$.
$A^i_j$, $B_i$, and $C_i$ are dimensionless constants.
$\delta m_{\hbox{\scriptsize I}}^{ij}$, $\delta\bar
m_{\hbox{\scriptsize I} k}^{ij}$, $a^i_j$, $A^i_j$, and $B_i$
give rise to LLI--violation, while $C_i$ and
$\delta m_{\hbox{\scriptsize P}ij}$ are responsible
for LPI--violation.
If all these coefficients vanish, we recover the usual
Schr\"odinger equation coupled to the Newtonian potential.
All coefficients are real.
It is clear that with the energy $m c^2$ and the
characteristic dimensionless quantity $U/c^2$ describing
a gravitational interaction, the GPE is the most general
2nd order differential equation including spin and the
gravitational potential tensor.
Only $U$, $U^{ij}$, $\phi_i$, $A_i$ and $H_{ij}$ are
$x$--dependent.
$H$ is hermitian.

\subsection{The classical limit}

The corresponding classical Hamilton function is 
\begin{eqnarray}
H^\prime & = & {1\over{2 m}} \left(\delta^{ij} + {{\delta m_{\hbox{\scriptsize I}}^{ij}}\over m} + 2 {{\delta\bar m_{\hbox{\scriptsize I} k}^{ij}}\over m} S^k\right) p_i p_j  - 2 \left({1\over m} a^i_j + c A^i_j\right) S^j p_i \nonumber\\
& & + e \phi + {e\over{m}} H_i S^i + 2 (m c^2 B_i + c T_i) S^i + (1 + 2 C_i S^i) m U + \delta m_{\hbox{\scriptsize P} ij} U^{ij}
\end{eqnarray}
where $p_i$ is the momentum and $S^i$ is the spin of the particle. 
The velocity, force and acceleration for vanishing electromagnetic field is 
\begin{eqnarray}
v^i & = & {1\over m} \left(\delta^{ij} + {{\delta m_{\hbox{\scriptsize I}}^{ij}}\over m} + 2 {{\delta\bar m_{\hbox{\scriptsize I} k}^{ij}}\over m} S^k\right) p_j - 2 \left({1\over m} a^i_j + c A^i_j\right) S^j \\
f_i & = & - (1 + 2 C_j S^j) m \partial_i U - \delta m_{\hbox{\scriptsize P} kl} \partial_i U^{kl} \\ 
a^i & = & - \left(\delta^{ij} + {{\delta m_{\hbox{\scriptsize I}}^{ij}}\over m} + 2 \left({{\delta\bar m_{\hbox{\scriptsize I} k}^{ij}}\over m} + \delta^{ij} C_k\right) S^k\right) \partial_j U - \delta^{ij} {{\delta m_{\hbox{\scriptsize P} kl}}\over m} \partial_j U^{kl} \label{accel}
\end{eqnarray}
where we neglected the dynamics of thr spin vector. 
This is reasonable because the corresponding interaction terms are very small. 
It is obvious that this generalises Haugan's result \cite{Haugan79} by
introducing the spin $S^k$.
In addition, a very important point which one can see by comparing (\ref{accel})
with (\ref{GPE}) is that in the GPE there are more LLI
and LPI--violating parameters than in the acceleration (\ref{accel}) of the corresponding classical particle with spin,
namely $A^i_j$ and $B_i$.
This acceleration is that quantity which is measured by E\"otv\"os--type experiments, like the torsion balance experiments, or the proposed experiments of the Bremen--tower and STEP. 
(Therefore (\ref{accel}) is a frame to describe tests of the equivalence principle for macroscopic matter with polarisation.)
That means: {\it If the classical acceleration fulfills LLI
and LPI then this does not rule out the possibility of
LLI and LPI--violating terms on the quantum level}.
{\it On the quantum level there are more possibilities to violate EEP than on the classical level}, even if one includes polarised bulk matter.

However, by considering the dynamics of spin, one gets access to all the EEP violating parameters. 
For the dynamics of the spin expectation value in the classical approximation we get
\begin{equation}
\frac{d}{dt} \mbox{\boldmath$S$} = \mbox{\boldmath$\Omega$} \times \mbox{\boldmath$S$} \label{SpinDynamics}
\end{equation}
with
\begin{equation}
\Omega_i := \frac{1}{2 m} \frac{\delta \bar m_{\hbox{\scriptsize I} i}^{kl}}{m} p_k p_l + \left(\frac{1}{m} a^j_i + c A^j_i\right) p_i + m c^2 B_i + c\, T_i + C_im U(x)
\end{equation}
That means that besides $\delta m_{\hbox{\scriptsize I}}^{ij}$ and $\delta m_{\hbox{\scriptsize P} kl}$ all the anomalous parameters influence the spin precession. 
In other words: Only if one takes the path (\ref{accel}) {\it and} the dynamics of the spin (\ref{SpinDynamics}) into account one can make statements about the complete set of parameters characterising the violation of EEP. 
However, the precession of the net polarisation of a macroscopic body is very difficult to observe. 
The corresponding quantum tests (see below) are much more sensitive. 

It is important that {\it any deviation from
the usual Schr\"odinger equation coupled to the Newtonian
potential} will give rise to LLI-- or LPI--violations.
Consequently, {\it a non-vanishing of one of the above parameters
implies that gravity is not describable by a Riemannian space-time metric.}

\subsection{Possible consequences of LLI violating parameters}

First we note that the coefficients $T$, $T_i$, and $a^i_j$ are due to the term $\Gamma$ which is not connected with the matrices $\widetilde\alpha^{\hat\mu}$ and $\widetilde\beta$ and therefore have no influence on the violation of LLI. 

Since we performed a non-relativistic limit we are not able to reduce the generalised Clifford algebra by means of possible experiments indicating that the coefficients $\delta m_{\hbox{\scriptsize I}}^{ij}$, $\delta \bar m_{\hbox{\scriptsize I} k}^{ij}$, $A^i_j$, $B_i$, and $C_i$ are zero. 
The only reasoning which can be done is that {\it if experiments indicate that one of the coefficients} $\delta m_{\hbox{\scriptsize I}}^{ij}$, $\delta \bar m_{\hbox{\scriptsize I} k}^{ij}$, $A^i_j$, $B_i$, {\it and $C_i$ is non--vanishing then at least one of the $X^{ij}$, $X^i$, and $X$ is non--vanishing also}.\footnote{Indeed, it is possible to show that (i) $B_i \neq 0 \Rightarrow X \neq 0$, (ii) $A^i_j \neq 0 \Rightarrow X^i \neq 0$, and  irreducible tensor part of $\delta m_{\hbox{\scriptsize I}}^{ij} \neq 0 \Rightarrow X^{ij} \neq 0$ (the trace results in a redefinition of the scalar mass $m$).} 
In other words: any deviation from the usual Schr\"odinger equation coupled to the Newtonian potential will give rise to a splitting of the null cone or of the mass shell, and in turn, to a LLI-- or LPI--violation. 

That means, a non-vanishing of the above parameters necessarily implies that there is no usual Clifford algebra. 
Therefore, there is no Riemannian metric which can be defined from the dynamics of the field under consideration and which is responsible for the dynamical behaviour of the quantum field. 
{\it A non-vanishing of one of the above parameters implies that gravity is not described by a Riemannian space---time metric.}

\section{Experimental restrictions}

\subsection{Matter wave interferometry}

We propose two different kinds of interference experiments
where we use an interferometer of Kaservich and Chu
type  \cite{KasevichChu91,KasevichChu92}.
The first is an experiment where only a spin-flip will be
performed whereby both parts of the matter wave propagate
with the same momentum.
The second is an interference experiment which
measures the acceleration.

\subsubsection{Spin-flip experiment}

We take an atomic beam with a
definite spin value along a certain axis propagating with
momentum $p_i$.
We split this atomic state into two states and perform
with one of these states two spin-flips, one at time $t$
and the second one reverses the first flip at time $t +
\Delta t$.
The phase shift after the second splin-flip (for convenience, we introduce $\hbar$ in an obvious way) $\phi =
{1\over\hbar} (H(p, S) - H(p, - S)) \Delta t$ is given by
(for $A_i = \phi = 0$)
\begin{equation}
\phi = {2\over\hbar} \left({{\delta\bar
m_{\hbox{\scriptsize I} k}^{ij}}\over{2 m^2}} p_i p_j -
{\hbar \over m} a^i_k p_i -
c A^i_k p_i + m c^2 B_k + C_k m U + c T_k\right) S^k \Delta t
\end{equation}
To first order we can replace the momentum by the
velocity,
and we use $U = G M/ r$.  Then
\begin{eqnarray}
\phi & = & {1\over\hbar} \left({{\delta\bar
m_{\hbox{\scriptsize I} k}^{ij}}\over m} p_i
\delta_{jk} l^k - 2 (c A^i_j m + a^i_j) \delta_{ik} l^k + 2 m c^2 B_j \Delta t + 2 C_j m
{{G M}\over R} \Delta t + \hbar c T_j \Delta t\right) S^j
\end{eqnarray}
For an atom interferometer of Kasevich and Chu type we have
$l \approx 1\;\hbox{cm}$, $m \approx 10^{-26}\;\hbox{kg}$,
$v \approx 10\;\hbox{cm/sec}$, $S = {1\over 2}$, and $\Delta t\approx 0.1\;\hbox{sec}$.
With the accuracy $\Delta\phi/\phi = 10^{-8}$ we can estimate
in the case that one performs a null-experiment:
$\left|\delta\bar m_{\hbox{\scriptsize I} k}^{ij}/ m\right|
\lesssim 10^{-7}$, $\left| A^i_j \right| \lesssim  10^{- 17}$, $|a^i_j| \lesssim 1\; {\hbox{m}}^{-1}$,
$\left| B_i \right| \lesssim 5\cdot 10^{-27}$, $\left| C_i \right|
\lesssim 10^{-17}$, and $|T_i| \lesssim 3\cdot 10^{-10}\,{\hbox{m}}^{-1}$.
For the first coefficient we may get a better estimate if we
take a large velocity $v = 10^3 \;\hbox{m}/\hbox{sec}$ and
$l = 100 \;\hbox{m}$.
We get $\left|\delta\bar m_{\hbox{\scriptsize I} k}^{ij}/m\right|
\lesssim 10^{-15}$.
This generalises results in \cite{ABL93} (see also \cite{Phillips87}).
If one of these quantities turns out to be non-null, then we
infer a violation of LLI and LPI.
However, all these quantities but the $A_j^i$ and $a^i_j$ can be measured better
by Hughes--Drever type experiments (see below).

\subsubsection{Measurement of acceleration}

For the atom beam interferometer of Kasevich and Chu \cite{KasevichChu91,KasevichChu92} we get a phase shift
$\phi = - k_i a^i T^2$ where $T$ is the time between the
laser pulses and $k_i$ is the difference of the wave
vectors of the two counterpropagating laser beams.
We use the acceleration (\ref{accel}) assuming that
the spin state is not affected by the beam splitting process.
For a spherical symmetric mass producing the gravitational
field with constant mass density we get the phase shift
($r_0^i$ is the vector
from the center of the gravitating body (say, the
earth) to the beam splitter)
\begin{eqnarray}
\phi & = & T^2
{{GM}\over{r_0^3}}\left(k_i r^i_0 +
{{\delta m_{\hbox{\scriptsize P} ij} }\over m} {{r_0^i
r_0^j}\over{r_0^2}}
k_l r^l_0 - {{6}\over
5}{{\delta m_{\hbox{\scriptsize P} ij}}\over m} r_0^i k^j \right. \nonumber\\
& & \left.  +
{2\over 5} {{\delta m_{\hbox{\scriptsize P} ii}}\over m}
k_l r^l_0 - {{\delta m_{\hbox{\scriptsize I} ij}(S)}\over m}
r_0^i k^j\right)
\label{KCphase}
\end{eqnarray}
where $\delta m_{\hbox{\scriptsize I} ij}(S)$ is the sum
of both anomalous inertial mass tensors which depend on the spin.
If we align $\mbox{\boldmath$x$}_0 \sim \vec e_z$
and denote the angle between $r_0^i$ and
$k^i$ by $\vartheta$ then we get
to first order in $\delta m_{\hbox{\scriptsize P} ij}$
\begin{equation}
\delta\phi = - \left(1 +
\alpha\right) k\, T^2 g \cos\left(\vartheta + b\right)
\label{KasevichChu}
\end{equation}
with
\begin{eqnarray}
\alpha & := & {1\over 5}\left({{\delta m_{\hbox{\scriptsize P}
zz}}\over m} + 4 {{\delta m_{\hbox{\scriptsize P} xx}}\over
m}\right) - {{\delta m_{\hbox{\scriptsize
I} zz}(S)}\over m} \\
b & := & {6\over 5} {{\delta m_{\hbox{\scriptsize
P} zx}}\over m} + {{\delta m_{\hbox{\scriptsize
I} zx}(S)}\over m} \label{alphab}
\end{eqnarray}
where $\delta m_{\hbox{\scriptsize P} xx} =
\delta m_{\hbox{\scriptsize P} yy}$ due to the spherical
symmetry of the gravitating body.
$\alpha$ and $b$ are constant for one sort of atoms.
Note that (\ref{KasevichChu}) is an exact quantum result
although there appears no $\hbar$.
The reason is that only experimentally given quantities
like $k$ and $T$ are used.
The non-diagonal parts $b$ induce a deviation from orthogonality:
By choosing $k_i$ orthogonal to $g^i \sim
r^i_0$, the phase shift will not vanish in
contrast to the case $b = 0$.
Instead we have to first order
\begin{equation}
\phi = b k T^2 g \, .
\label{KCb}
\end{equation}
The phase shift will vanish only for an angle $\vartheta = -
(6/5)\delta m_{\hbox{\scriptsize P} zx}/m -
\delta m_{\hbox{\scriptsize
I} zx}(S)/m$.
The diagonal parts $\alpha$ modify the amount of the phase shift for
given $k$ and flight time $T$ and are best measured
if $k^i \sim g^i$.
The phase shift is to first order
\begin{equation}
\phi = - (1 + \alpha) k T^2 g
\, . \label{KCalpha}
\end{equation}
In order to measure $\alpha$ and $b$ we consider two cases: (i)
$\vartheta = 0$ with the phase shift (\ref{KCalpha}) for
measuring $\alpha$ and (ii) $\vartheta = \pi/2$ with the phase
shift (\ref{KCb}) for measuring $b$ where we use
$g := \nabla U = G M_\oplus/r_0^2$
as acceleration of the earth's gravitational field at the
position of the beam splitter.

A possible way to measure $\alpha$ is to consider the variation
of the phase shift (\ref{KCalpha}) by varying the elevation of
the whole interferometer with respect to the surface of the
earth, that is, by varying $r$ from $r_0 = R_\oplus$ to $r_1 =
r_0 + \delta r$.
This procedure is especially appropriate for the Raman light-
pulse atom beam interferometer because it may be possible to
detect by means of phase shifts the difference in the elevation
of 1mm.
This corresponds to an accuracy of the measured phase shift of
$\Delta\phi/\phi \approx 10^{-10}$ which \cite{KasevichChu92}
claim to be possible to achieve.
The today's accuracy is $10^{-8}$.
Using this experimental setup one gets estimates for $\alpha$
by detecting deviations of the measured phase shift from
$\displaystyle \delta\phi_0 =  k T^2\Delta g\cos\vartheta$
where $\Delta g = g(R_\oplus + \delta r) - g(R_\oplus)$ is the
difference of the acceleration between $R_\oplus + \delta r$
and $R_\oplus$.
Thereby the acceleration or the earth's gravitational potential
is measured by gravimeters (or satellites) which are of
other material composition than the interfering atoms.
Therefore we actually compare the possible composition
dependence of the gravitational acceleration between the atom
beam and other gravimeters.

In principle the energy levels of the atoms are also
modified due to a violation of LPI and LLI (see
\cite{Will74,GabrielHaugan90} for calculations within the $TH\epsilon\mu$--formalism and below).
However, this does not matter since the wave vector and frequency
of the laser beam which are
related to the difference of the energy levels are given
quantities in the quasi Newtonian coordinate system.
Consequently, we can use formula (\ref{KasevichChu}) with the
actually used momentum $k$ to determine $\alpha$:
$1 + \alpha = \phi/k\,
T^2 \Delta g$ where
$\phi$ is the measured
variation of the phase shift
during elevation of the interferometer from $r_0$ to $r_0 +
\delta r$.
If one assumes a null experiment, then the
accuracies of the various entities entering the phase shift
will give an estimate of the value of $\alpha$
which depends on the accuracy of
$\phi$, $k T^2$, and $\Delta g$.
The first two have been estimated in
\cite{KasevichChu91,KasevichChu92} to $3\cdot 10^{-8}$ and
the best absolute gravimeters have an accuracy of $10^{-6}$.
Consequently, an observation of the phase shift by elevating
the whole interferometer about $\delta r = 100$m and a null
result will give $|\alpha| \lesssim
3\cdot 10^{-6}$ and consequently $\displaystyle \left|{{\delta
m_{\hbox{\scriptsize P} xx}}\over m} + 4 {{\delta
m_{\hbox{\scriptsize P} zz}}\over m} - 5 {{\delta
m_{\hbox{\scriptsize I} zz}(S)}\over m}\right|
\lesssim 1.5\cdot 10^{-5}$.
With the results of Hughes--Drever type
experiments (see below)
these experiments amount to an estimate
$|\delta m_{\hbox{\scriptsize P} ij}/m| \lesssim 10^{-7}$
which is usually tested by red--shift experiments to $\lesssim 10^{-4}$,
see \cite{Will93,Godoneetal95}, for example.

A second way to measure this effect is to take a second
apparatus with a different kind of atoms, elevate the two
apparatus, and to compare the phase shifts.
Here one does not need a measurement of the gravitational
acceleration by some gravimeter.
In this case one measures the difference $\alpha_1 - \alpha_2$
with, for a null result, an accuracy given by the atom beam
interferometers $|\alpha_1 - \alpha_2| \lesssim 6\cdot 10^{-8}$.

A third way to measure $\alpha$ with (\ref{KCalpha}) is
to stay with
the apparatus at the same position and vary $T$ which leads
approximately to the same estimate for $\alpha$ and $\delta
m_{\hbox{\scriptsize P}}/m$ as above.

For measuring $b$ we have to align $k_i$
orthogonal to $g^i$.
For a given initial velocity of the atoms and a specially
chosen time $T$ one can arrange that the atoms may fly parts of
parabolae so that the $\pi/2$ pulses interact with the beams on
the same height  and the $\pi$ pulse hits both beams at their
peak.
If the experiment gives a null result then the accuracy
of $b$ is limited by the accuracy of fixing
$k_i$ exactly orthogonal to $g^i$.
Since this accuracy is approximately $10^{-6}$ (again given by
satellite results) we have $|b| \lesssim 10^{-6}$ and
consequently $|\delta m_{\hbox{\scriptsize P} zx}/m + (5/6) \delta m_{\hbox{\scriptsize I} zx}(S)/m| \lesssim 10^{-6}$.

\subsection{Hughes-Drever experiment \label{HughesDrever}}

We can also use the above Hamiltonian to calculate the
Zeeman--splitting of energy levels in an atom.
In order to do that we first have to consider a two-particle
system and to eliminate the center--of--mass motion.
In the case of a heavy nucleus, we can neglect the
difference between the mass and the reduced mass.
With the single particle Hamiltonian (\ref{GPE}) and
fixing the nucleus at $x = 0$ we get the Hamiltonian
describing the energy levels of an electromagnetically bound system:
\begin{equation}
H_{\hbox{\scriptsize I}} = H_0 + H_{\hbox{\scriptsize I, em}} + H_{\hbox{\scriptsize I, non-Einst.}}  \label{Ham}
\end{equation}
with
\begin{eqnarray}
H_0 & = & - {{\hbar^2}\over{2 m}} \Delta + {{Z e^2}\over{|x|}} + V_{\hbox{\scriptsize nucl}}(x) \\
H_{\hbox{\scriptsize I, em}} & = & {e\over{2 m}} H_i (L^i + \sigma^i)  \\
H_{\hbox{\scriptsize I, non-Einst.}} & = & - {{\hbar^2}\over{2 m}}
{{\delta m_{\hbox{\scriptsize I}}^{ij}(\sigma)}\over{m}}
{\partial\over{\partial x^i}} {\partial\over{\partial x^j}}
+ \left({1\over m} a^i_j + c A^i_j\right) \sigma^j i \hbar {\partial\over{\partial x^i}} - \hbar c T_i \sigma^i + c^2 m B_i \sigma^i \nonumber\\
& & + m C_i \sigma^i U - \delta m_{\hbox{\scriptsize I}}^{ij}(\sigma) x_i \nabla_j U
- \delta m_{P kl} x^i \nabla_i U^{kl} \label{HnonEinst}
\end{eqnarray}
and where $V_{\hbox{\scriptsize nucl}}$ is the nuclear
binding potential.
Here $x$ is the relative coordinate between e.g.\ the core of a
nucleus and a valence proton.
Note that there are no Einsteinian effects due to the
acceleration $\nabla U$.
This is in accordance with the equivalence principle:
The effect of gravitational acceleration can be cancelled
by a transformation to a suitable accelerated frame.
$H_0$ describes the atom without external fields and without
disturbances, $H_{\hbox{\scriptsize I, em}}$ is the interaction
of the electron with an external constant magnetic field, and $H_{\hbox{\scriptsize I, non-Einst.}}$ gives LLI and
LPI-violating effects.
Terms linear in $x^i$ do not contribute in first order to
energy shifts.
(\ref{Ham}) is the equation for the energy levels.

We can consider the case of an atomic nucleus which
consists of a core and a valence proton which is considered
in the usual Hughes-Drever experiments (see e.g. \cite{Will93}).
In the case of ${}^7\hbox{Li}$ we have a $J = 0$ core and a
valence proton with angular momentum $L = 1$ and spin $1\over 2$.
In \cite{Lamoreauxetal86} and \cite{Chuppetal89}
${}^{201}{\hbox{Hg}}$ and ${}^{21}{\hbox{Ne}}$ with
a similar nuclear structure was used.
With the wave functions $|J, M_J\rangle$ for the magnetic
quantum numbers $M_J = \left\{{3\over 2},
{1\over 2}, - {1\over 2},- {3\over 2}\right\}$ the matrix
elements of $H_{\hbox{\scriptsize I, non-Einst.}}$ are
easily calculated whereby the matrix elements linear
in momentum and in the relative position vanish.
Therefore, the energy levels for the hyperfine-splitting
are shifted and the singlet line thus splits into
three lines with the energies
\begin{eqnarray}
E({\scriptstyle {3\over 2}, {3\over 2}}) -
E({\scriptstyle {3\over 2}, {1\over 2}}) & = & \delta +
\bar\delta_1 + \bar\delta_2  \\
E({\scriptstyle {3\over 2}, {1\over 2}}) -
E({\scriptstyle {3\over 2}, - {1\over 2}}) & = &
\bar\delta_2 \\
E({\scriptstyle {3\over 2}, - {1\over 2}}) -
E({\scriptstyle {3\over 2}, - {3\over 2}}) & = & - \delta +
\bar\delta_1 + \bar\delta_2
\end{eqnarray}
with
\begin{eqnarray}
\delta & := & - {{\hbar^2}\over{3 a^2}} \left(
{{\delta m_{\hbox{\scriptsize I}}^{xx}}\over{m^2}} +
{{\delta m_{\hbox{\scriptsize I}}^{yy}}\over{m^2}} - 2
{{\delta m_{\hbox{\scriptsize I}}^{zz}}\over{m^2}}\right) \\
\bar\delta_1 & := & - {{\hbar^2}\over{a^2}} \left({{\delta
\bar m_{\hbox{\scriptsize I}z}^{xx}}\over{m^2}} +
{{\delta \bar m_{\hbox{\scriptsize I}z}^{yy}}\over{m^2}} -
2 {{\delta \bar m_{\hbox{\scriptsize I}z}^{zz}}\over{m^2}}\right)
\\
\bar\delta_2 & : = & - {{5\hbar^2}\over{3 a^2}}
{{\delta \bar m_{\hbox{\scriptsize I}z}^{zz}}\over{m^2}} +
{2\over 3} (m c^2 B_z + m C_z U) + c \hbar T_z
\end{eqnarray}
where we modelled $V_{\hbox{\scriptsize nucl}}$ by the harmonic
oscillator potential.
$a$ is the radius of the nucleus.
$\delta$ and $\bar\delta_1$ is responsible for a splitting
of the single line into three lines, and $\bar\delta_2$
shifts all three lines in the same way.
The search for these splittings during the change
of the $z$-axis with respect to the nonrotating
Newtonian coordiante system amounts to a test of LLI
and LPI--violation.
Using the experimental accuracy of this type of experiment
(see \cite{Chuppetal89}, also
\cite{Prestageetal85,Lamoreauxetal86}) we have
$\delta E < 0.45\cdot 10^{-6}\,\hbox{Hz}$ so that,
provided no unfortunate cancellation of terms occurs,
we get the estimates  
$|\delta\bar m_{\hbox{\scriptsize I}z}^{xx}/m|
\lesssim 5\cdot 10^{-30}$, $|B_z| \lesssim 3\cdot 10^{-31}$,
$|C_z|
\lesssim 3\cdot 10^{-24}$, and $|T_z| \lesssim 1.5\cdot 10^{-15}\,{\hbox{m}}^{-1}$ where we used $a = 1.5\;\hbox{fm}$
and the gravitational
potential
$U/c^2 \approx 10^{-7}$ of our galaxy.
The estimate on $T_i$ may be interpreted as the till now tightest constraint on a hypothetical axial torsion \cite{La97}. 

Since only the relative coordinate
$x$ appears in (\ref{HnonEinst})
our cause of violation of LLI is different from
that described by the $TH\epsilon\mu$-formalism
\cite{LightmanLee73,Will93}:
While in the latter case  the LLI-violation
is due to a relative velocity with respect to the
rest frame of the universe, it is in our case due
to the anomalous mass tensors appearing
in the GPE.
These LLI-violating terms cannot be transformed
away by choosing a special frame of reference.

\section{Conclusions and Discussion}

We have presented a general approach to a Pauli equation describing all possible anomalous couplings of the matter field to the Newtonian gravitational field and to hypothetical other gravitational fields. 
In deriving this equation we did not use any theoretical concept which has no direct physical interpretation nor any geometrical notion. 
The latter should be a consequence of experiments. 
Therefore we obtained in a systematic manner all anomalous couplings on the non-relativistic level, thus enlarging in a non--trival way Haugan's test theory. 
Main features of our quantum test theory are that we include spin in a non--trivial way and that it possesses more possibilities to violate LLI and LPI than the corresponding classical theory.

We want to remark that the $TH\epsilon\mu$--formalism leads to anomalous coupling terms which are part of our set of anomalous terms derived above.
Consequently, if one of these parameters does not vanish, then it needs further experimental and theoretical analysis in order to distinguish whether this anomalous term is due to an underlying GDE considered in this work, or due to an anomalous coupling between the quantum field and the Maxwell field.

\section*{Acknowledgement}

I thank Ch.J.\ Bord\'e and R.\ Kerner for the hospitality at the Laboratoire de Gravitation et Cosmologie Relativistes, at the Universit\'e Perre et Marie Curie, Paris, and the Deutsche Forschungsgemeinschaft for financial support.

\end{document}